# HEART RATE VARIABILITY MONITORING IDENTIFIES ASYMPTOMATIC TODDLERS EXPOSED TO ZIKA VIRUS DURING PREGNANCY

Christophe L. Herry [1], Helena M.F. Soares [2], Lavinia Schuler-Faccini [2], Martin G. Frasch [3]

[1] Ottawa Hospital Research Institute, University of Ottawa, ON, Canada
[2] INAGEMP - Departamento de Genética – Instituto de Biociências, Universidade Federal do Rio Grande do Sul, Brazil
[3] Department of Obstetrics and Gynecology, University of Washington, Seattle, WA, USA

Address of correspondence:
Dr. Martin G. Frasch
Department of Obstetrics and Gynecology
University of Washington
1959 NE Pacific St
Box 356460
Seattle, WA 98195
Phone: +1-206-543-5892
Fax: +1-206-543-3915
Email: mfrasch@uw.edu

**Word count:** 1551

**Key Points**

**Question**

Can single-channel electrocardiogram (ECG) - derived heart rate variability (HRV) recorded in toddlers yield a biomarker of fetal in utero exposure to Zika virus?

**Findings**

Grid Count measure of HRV serves as the best ECG-derived biomarker of fetal exposure to Zika virus infection with high classification performance.

**Meaning**

It is feasible to measure HRV in infants and toddlers using a small non-invasive portable ECG device. Such screening approach may uncover memory of *in utero* exposure to ZIKV.

**Abstract**

Although Zika virus (ZIKV) seems to be prominently neurotropic, there are some reports of involvement of other organs, particularly the heart. Of special concern are those children exposed prenatally to ZIKV and born with no microcephaly or other congenital anomaly. Electrocardiogram (ECG) - derived heart rate variability (HRV) metrics represent an attractive, low cost, widely deployable tool for early identification of such children. We hypothesized that HRV in such children would yield a biomarker of fetal ZIKV exposure. We investigated the HRV properties of 21 infants aged 4 to 25 months from Brazil. The infants were divided in two groups, the ZIKV-exposed (n=13) and controls (n=8). Single channel ECG was recorded in each child at ~15 months of age and HRV was analyzed in 5 min segments to provide a comprehensive characterization of the degree of variability and complexity of the heart rate. Using a cubic Support Vector Machine (SVM) classifier we identified babies as Zika cases or controls with negative predictive value of 92% and positive predictive value of 86%. Our results show that HRV metrics can help differentiate between ZIKV-affected, yet asymptomatic, and non-ZIKV exposed babies. We identified the Grid Count as the best HRV measure in this study allowing such differentiation, regardless the presence of microcephaly. We show that it is feasible to measure HRV in infants and toddlers using a small non-invasive portable ECG device and that such approach may uncover memory of *in utero* exposure to ZIKV. This approach may be useful for future studies and low-cost screening tools involving this challenging to examine population.

**Introduction**

The sudden outbreak of children being born with microcephaly in Brazil in 2015 led to the identification of a new teratogen in humans: the Zika Virus (ZIKV).[1,2] A specific pattern of "Fetal Brain Disruption Sequence" [3] with small skull and pronounced craniofacial disproportion, overlapping cranial sutures, occipital prominence and redundant skin in the head and neck, joint contractures, and severe neurological impairments was described, and brain imaging (CT scan or MRI) presenting multiple calcifications, poor and abnormal gyral patterns, increased intracranial fluid, and cortical destruction. [4,5] However, there is a spectrum of less severe abnormalities after congenital ZIKV exposure with a range of abnormalities including postnatal microcephaly, hydrocephaly, brain calcifications, ophthalmic abnormalities, seizures and other neurological dysfunctions.[6–8]

Although ZIKV seems to be prominently neurotropic, there are some reports of involvement of other organs, particularly the heart. Cavalcanti et al. found 14/103 (13.5%) Brazilian children with congenital zika syndrome (CZS) to present with echocardiograms compatible with congenital heart disease.[9] Most of them, however, were septal defects with no hemodynamic significance. Similarly, Orofino et al. reported 13/97 (10.8%) major cardiac defects (atrial septal defect, ventricular septal defect, patent ductus arteriosus) on echocardiographic exams in infants with confirmed ZIKV exposure.[10] None of the defects were severe.

Of special concern are those children exposed prenatally to ZIKV born with no microcephaly or other congenital anomaly at birth. Electrocardiogram (ECG) - derived heart rate variability (HRV) metrics represent an attractive, low cost, widely deployable tool for early identification of such children.

We hypothesized that ECG-derived HRV in such children would yield a biomarker of fetal ZIKV exposure.

**Methods**

We investigated the HRV properties of 21 infants aged 4 to 25 months from Brazil. The study was approved by the local Ethics Board (CAAE: 56176616.2.1001.5327). This was a convenience sample from a group of children being followed by us according to the guidelines of the Surveillance and Response Protocol published by the Brazilian Ministry of Health.[11] All participants provided full written informed consent for data collection and registration.

Congenital infections were diagnosed using the following criteria: 1) ZIKV: confirmed by maternal polymerase chain reaction (PCR) result (in blood) consistent with Zika virus disease during pregnancy;.2) CMV: confirmed by positive PCR in urine of the baby.

The infants were divided in two groups, the ZIKV-exposed and controls: (1) Group with ZIKV maternal infection during pregnancy (n=13) and (2) a control group (n=8) of children with other diagnoses reported to authorities due to microcephaly (4 congenital infection, 1 genetic syndrome, 1 small for gestational age – normal development; 2 normal infants). In the exposed group, two were diagnosed with CZS with severe microcephaly. The remaining 11 children were born with normal head circumference (HC), but to mothers with RT-PCR confirmed ZIKV infection during pregnancy (n=9) or probable ZIKV infection by clinical/epidemiological criteria (n=2). Symptoms of ZIKV in this sample were in the first trimester of pregnancy (n=3); second (n=6) and third (n=4).

Single channel ECG was recorded in each child at ~15 months of age using the Zephyr device (Zephyr Technology Corporation, Annapolis, MD) for 20 minutes on average. Recording was performed in the morning, before the clinical examination with the baby held by the mother in the upright position. Baby/infant was calm and awake. We excluded children who got too agitated or with excessive crying.

HRV analysis: R-peaks were extracted automatically from the ECG recordings in order to create R-peak to R-peak interval time series. The R-peak locations were visually inspected to make sure the detection algorithm worked correctly. Noisy portions (saturated, very noisy periods) were discarded from the subsequent analysis.

On average, around 15 minutes of ECG recordings were usable for the HRV analysis.

The HRV analysis was done using 5 min moving windows and 2.5 min overlap and a set of 54 variability metrics were calculated for each 5 min window, in order to provide a comprehensive characterization of the degree of variability and complexity of the heart rate.[12,13] Each variability metric time series was then averaged over all 5 min windows.

Classification: A cubic Support Vector Machine (SVM) classifier was used to classify babies as Zika cases or controls. The SVM kernel was scaled automatically and we used a regularization factor of 1. A 100 repeat, 5-fold cross-validation process was used to select the best single variability metric and estimate the performance the cubic SVM classifier on the 20 subjects to estimate classification performance. A single metric was selected due to the limited sample size.

Both HRV processing and classification were implemented in Matlab R2016b (Mathworks Inc.).

**Results**

The clinical characteristics of the cohort are summarized in the Table 1. As expected, the number of cases with microcephaly, but not head circumference, was lower in the ZIKV group compared to controls. There were more younger infants in the ZIKV group during the ECG recording than in the control group. Other characteristics were similar.

The best HRV metric was the Grid Count, a complexity metric derived from a grid transformation of a dynamical system trajectory.[14] Results in the following Table 2 are averaged over all repeats/folds of the cross-validation process. Fig. 1 shows that Grid Count measure of HRV for these two babies was within the range of other ZIKV-group subjects. Excluding the two ZIKV-babies with microcephaly did not change these findings.

**Discussion**

Our results show that HRV metrics can help differentiate between ZIKV-affected, yet asymptomatic, and non-ZIKV exposed babies. We identified the Grid Count as the best HRV measure in this study allowing such differentiation, regardless the presence of microcephaly. Future studies including a larger sample size will permit to possibly identify an even more precise subset of HRV measures with regard to ability to differentiate between ZIKV-exposed babies and controls.

Our finding aligns with the recently published study in *ex vivo* fetal sheep hearts where we identified this measure as correlated to left ventricular end-diastolic pressure. [15]

ZIKV exposure *in utero* causes placental hypoperfusion.[17] Understanding that ZIKV exposure can create chronic hypoxic environment for the fetus, we can attempt to explain our HRV finding.

First, albeit temporal profile of Grid Count measure has not been studied in human development yet, we can expect with a reasonable confidence based on the behavior of other HRV metrics [18] that the overall increase of the gcount metric seen in Figure 1 reflects system maturation, i.e., increase in complexity.

Second and most intriguingly, the relationship between ZIKV-exposed and controls is inverse compared to what we see *ex vivo*. In the setting of chronic fetal hypoxia the Grid Count appears to serve as memory of *in utero* hypoxia in the iHRV *ex vivo*. *Ex vivo*, a removal of hypoxically triggered chronic sympathetic hyperactivity may manifest as a rise in Grid Count. Increasing beta-adrenergic drive on cardiac pacemaker cells synchronizes their activity in acute mechanistic experiments.[19] These findings likely hold true for the synchronizing impact chronic sympathetic activity exerts on sinus node pacemaker cell coupling *in vivo*. A withdrawal of such driving input might be expected to result in a rebound of uncoupling which would explain rise in Grid Count.

Conversely, its developmental counterpart *in vivo* would behave in the opposite, because a postnatal intact innervated heart does not experience the rebound response we observe *ex vivo*. Along such line of thought, chronic hypoxia with elevated sympathetic activity would continue on postnatally imprinting stronger cardiac pacemaker cell coupling and relatively lower Grid Count.

While the present results are encouraging and classification performance is surprisingly high, these findings remain very preliminary and limited due to the small sample size and will need to be confirmed and expanded on a much larger cohort.

A limitation of this pilot study is that there was some variance in ages during ECG recording which may have contributed to the difference in HRV measures, because, as discussed above, HRV measures may show maturational increase. A counter-argument to a confounding effect of HRV maturation is that despite more cases of microcephaly, head circumference was similar in both groups.

Another limitation arises from the wide range of gestation when the presence of ZIKV symptoms was reported. Future studies will need to show whether such broad developmental window of exposure has a variable effect on the HRV phenotype. The more interesting is the impact on one particular HRV measure, Grid Count, despite the wide developmental window of exposure. Considering that the pacemaker cell activity commences as early as 24 days of gestation[20], even early Zika exposure may result in similar effects we capture compared to late exposure. This is a speculation based on the current pilot study and warrants further investigations into how exposure to the virus during different trimesters impacts HRV.

Lastly, two of the ZIKV-exposed babies presented with a microcephaly, i.e., unlike the remaining eleven subjects, were clinically symptomatic. We confirmed that these two subjects had however no significant impact on our HRV findings.

In summary, here we show that it is feasible to measure HRV in infants and toddlers using a small non-invasive portable ECG device and that such approach may uncover memory of *in utero* exposure to ZIKV. This approach may be useful for future studies involving this challenging to examine population.

**Acknowledgements**: Funded by the University of Washington Global Innovation Fund (UW GIF) (LSF, CH, MGF); Canadian Institutes of Health Research (MGF); Brazilian Ministry of Health, and by CNPq – Brazilian National Council of Research and Technology (LSF).

**References**

1. Schuler-Faccini L, Sanseverino M, Vianna F, et al. Zika virus: A new human teratogen? Implications for women of reproductive age. *Clin Pharmacol Ther*. 2016;100(1):28-30.

2. Rasmussen SA, Jamieson DJ, Honein MA, Petersen LR. Zika Virus and Birth Defects--Reviewing the Evidence for Causality. *N Engl J Med*. 2016;374(20):1981-1987.

3. Russell LJ, Weaver DD, Bull MJ, Weinbaum M. In utero brain destruction resulting in collapse of the fetal skull, microcephaly, scalp rugae, and neurologic impairment: the fetal brain disruption sequence. *Am J Med Genet*. 1984;17(2):509-521.

4. Moore CA, Staples JE, Dobyns WB, et al. Characterizing the Pattern of Anomalies in Congenital Zika Syndrome for Pediatric Clinicians. *JAMA Pediatr*. 2017;171(3):288-295.

5. Del Campo M, Feitosa IML, Ribeiro EM, et al. The phenotypic spectrum of congenital Zika syndrome. *Am J Med Genet A*. 2017;173(4):841-857.

6. Aragao MFVV, Holanda AC, Brainer-Lima AM, et al. Nonmicrocephalic Infants with Congenital Zika Syndrome Suspected Only after Neuroimaging Evaluation Compared with Those with Microcephaly at Birth and Postnatally: How Large Is the Zika Virus "Iceberg"? *AJNR Am J Neuroradiol*. 2017;38(7):1427-1434.

7. van der Linden V, Pessoa A, Dobyns W, et al. Description of 13 Infants Born During October 2015-January 2016 With Congenital Zika Virus Infection Without Microcephaly at Birth - Brazil. *MMWR Morb Mortal Wkly Rep*. 2016;65(47):1343-1348.

8. Zin AA, Tsui I, Rossetto J, et al. Screening Criteria for Ophthalmic Manifestations of Congenital Zika Virus Infection. *JAMA Pediatr*. 2017;171(9):847-854.

9. Di Cavalcanti D, Alves LV, Furtado GJ, et al. Echocardiographic findings in infants with presumed congenital Zika syndrome: Retrospective case series study. *PLoS One*. 2017;12(4):e0175065.

10. Orofino DHG, Passos SRL, de Oliveira RVC, et al. Cardiac findings in infants with in utero exposure to Zika virus- a cross sectional study. *PLoS Negl Trop Dis*. 2018;12(3):e0006362.

11. *Protocolo de Vigilância E Resposta a Ocorrência de Microcefalia E/ou Alterações Do Sistema Nervoso Central (SNC)*. Brazilian Ministry of Health; 2016. http://combateaedes.saude.gov.br/images/sala-de-situacao/Microcefalia-Protocolo-de-vigilancia-e-resposta-10mar2016-18h.pdf.

12. Durosier LD, Herry CL, Cortes M, et al. Does heart rate variability reflect the systemic inflammatory response in a fetal sheep model of lipopolysaccharide-induced sepsis? *Physiol Meas*. 2015;36(10):2089-2102.

13. Bravi A, Longtin A, Seely AJE. Review and classification of variability analysis techniques with clinical applications. *Biomed Eng Online*. 2011;10:90.

14. Roopaei M, Boostani R, Sarvestani RR, Taghavi MA, Azimifar Z. Chaotic based reconstructed phase space features for detecting ventricular fibrillation. *Biomed*

*Signal Process Control*. 2010;5(4):318-327.

15. Frasch MG, Herry C, Niu Y, Giussani DA. First evidence that intrinsic fetal heart rate variability exists and is affected by chronic hypoxia. *bioRxiv*. August 2018:242107. doi:10.1101/242107

16. Frasch MG, Herry CL, Niu Y, Giussani DA. First Evidence of Intrinsic Fetal Heart Rate Variability Affected by Chronic Fetal Hypoxia. In: *REPRODUCTIVE SCIENCES*. Vol 24. SAGE PUBLICATIONS INC 2455 TELLER RD, THOUSAND OAKS, CA 91320 USA; 2017:198A - 198A.

17. Hirsch AJ, Roberts VHJ, Grigsby PL, et al. Zika virus infection in pregnant rhesus macaques causes placental dysfunction and immunopathology. *Nat Commun*. 2018;9(1):263.

18. Van Leeuwen P, Lange S, Bettermann H, Gronemeyer D, Hatzmann W. Fetal heart rate variability and complexity in the course of pregnancy. *Early Hum Dev*. 1999;54(3):259-269.

19. Yaniv Y, Ahmet I, Liu J, et al. Synchronization of sinoatrial node pacemaker cell clocks and its autonomic modulation impart complexity to heart beating intervals. *Heart Rhythm*. 2014;11(7):1210-1219.

20. Weisbrod D, Khun SH, Bueno H, Peretz A, Attali B. Mechanisms underlying the cardiac pacemaker: the role of SK4 calcium-activated potassium channels. *Acta Pharmacol Sin*. 2016;37(1):82-97.

**Table 1: Clinical features among cases and controls.**

| | CASES (TOTAL n = 8) | CONTROLS (TOTAL n = 13) | p * |
|---|---|---|---|
| | N% | N% | |
| Sex Masculine | 7 (87.5%) | 7 (53.8%) | 0.173 |
| Microcephaly | 7 (87.5%) | 2 (15.3%) | 0.002 |
| **Age At Heart Exam** | | | |
| 4-12 Months | 5 (62.5%) | 0 | 0.014 |
| 13-24 Months | 3 (37.5%) | 13 (100%) | |
| **Gestational Age** | | | |
| ≤ 36 Weeks | 2 (25%) | 2 (15.3%) | 0.498 |
| ≥ 37 Weeks | 6 (75%) | 11 (84.6%) | |
| **Head Circumference Z-score** | | | |
| < -3 | 3 (37.5%) | 2 (15.3%) | 0.120 |
| -3 to -2 | 1 (12.5%) | 0 | |
| ≥ -2 | 4 (50%) | 11 (84.6%) | |
| **Zikv Symptoms** | | | |
| 1st Trimester | - | 3 (23.0%) | |
| 2nd Trimester | - | 6 (46.1%) | |
| 3rd Trimester | - | 4 (30.7%) | |

* Fisher Exact Test

**Table 2. Identification of ZIKV exposed / control toddlers using HRV metric Grid Count.**

| AUC | SEN | SPE | NPV | PPV |
|---|---|---|---|---|
| 94.51% | 85.71% | 92.31% | 92.31% | 85.71% |

AUC: Area under the Receiver Operating Curve; SEN: sensitivity; SPE: specificity; NPV: negative predictive value; PPV: positive predictive value.

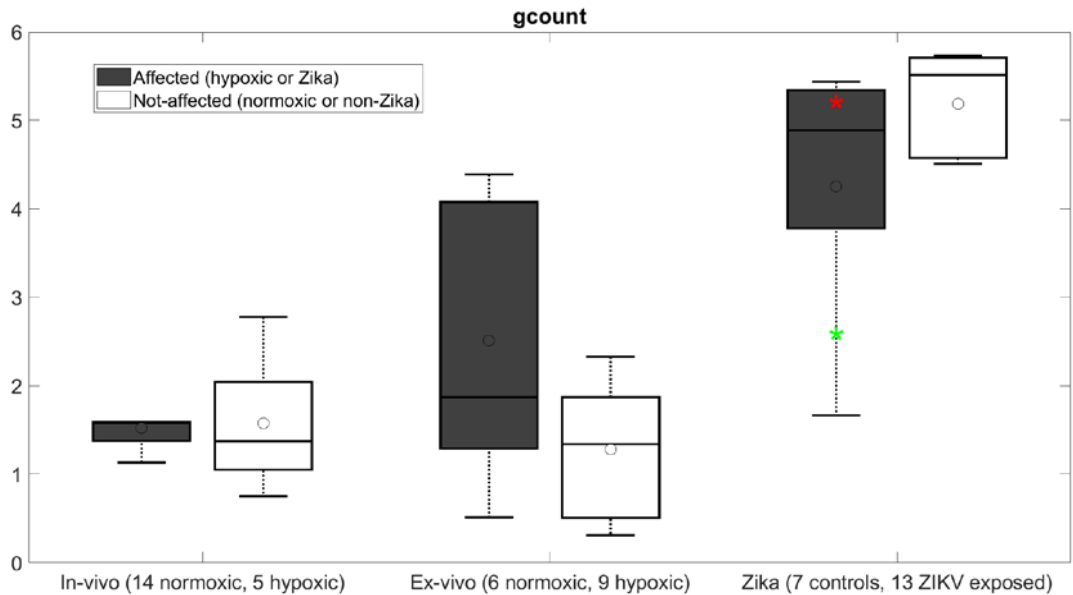

**Figure 1.** The discriminating HRV measure Grid Count (gcount) is shown in hypoxic versus normoxic (both *in-vivo* and *ex-vivo*) near-term ovine fetuses (presented elsewhere [15,16]) and ZIKV-exposed versus control human cohorts. The rationale for this juxtaposition is the purported chronic hypoxic effect of ZIKV exposure on fetal development. See discussion for details.

Horizontal lines are median and open circles are mean. The box is the interquartile range. Outliers are not shown but "whiskers" represent 95%. Red and green * represent the two cases of microcephaly in the ZIKV-exposed group. ZIKV versus control: p=0.096; ZIKV vs in-vivo hypoxic: p<0.001; Zika control vs in-vivo normoxic: p<0.001 (All tests are Wilcoxon rank sum tests).